# LIBRARY INFORMATION SYSTEM AUDIT SENAYAN LIBRARY MANAGEMENT SYSTEM (SLIMS) USING ISO 9126

**Petrus Dwi Ananto Pamungkas**

Department of Informatics Engineering; STMIK Bina Insani;
e-mail: petrusdwi@binainsani.ac.id

## Abstract

The library serves as a vehicle for education, research, conservation, information, and recreation to improve the nation's intelligence and empowerment [1]. The function of the library as a place of education, research, and information provides an opportunity to use the information system of Senayan Library Management System (SLiMS) in the library in order to improve the service to the user, increase the reading interest, and expand the insight and knowledge to educate the nation. The use of ISO 9126 standard is able to know the quality of SLiMS information system which is said to be free of charge of usage and license (because it belongs to Open Source Software category [2]) to assist library management in Indonesia. The implementation of the SLiMS information system audit in several university libraries refers to the ISO 9126 standard by using the Functionality, Reliability, Usability, Efficiency, Maintainability and Portability aspects through distributing questionnaires to university librarians in charge. With the help of the use of Google Forms it turns out that only ten universities librarians in charge who are willing to fill out the questionnaires are IPMI IBS, Bakrie University, Perbanas Institute Jakarta, STMIK & Bina Insani Academy, Prasetya Mulya University, Agung Podomoro University, Indonesian Higher Law School, Matana University, STIKS Tarakanita Jakarta, and STAI-PIQ West Sumatra. From the results of data processing it is known that SLiMS included in the category VERY GOOD for use in the management of libraries in college. This means that the ten universities librarians in charge admitted and have proven that SLiMS is very helpful in library management.

**Keywords**: library, Senayan Library Management System (SLiMS), audit information system, ISO 9126.

## A. Introduction

Library serves as a vehicle for education, research, conservation, information, and recreation to improve the intelligence and empowerment of the nation. The library aims to provide services to the user, to increase the passion of reading, and to broaden the insight and knowledge to improve the life of the nation [1]. The college library develops library services based on information and communication technologies.

The use of information systems in the implementation must be in accordance with the direction of library management objectives that exist in each college. Library management can include planning, organizing, actuating and controlling. Because if any sophisticated information system has been created but not in accordance with the college library management, it will be less than optimal use of the information system. For example, the existing library system has a less familiar display so that the librarian takes a long time to understand its use and the librarian in charge sometimes gets into trouble when integrating data in the form of other files (Ms. Word, Ms. Excel, and so on).

The presence of information system Senayan Library Management System (SLiMS) which is the work of the Indonesian nation is expected to become a reliable library managing application. Its use is also exempt from usage fees or license fees as it is included in the Open Source Software category [2]. To know the quality of information systems SLiMS then conducted an audit information system using ISO 9216 standard which is one of the international standards are widely used.

## B. Literature Review

1.  Basic Concept of Information System Auditing

The information system is a collection of interconnected elements that form a unity to integrate data, process and store also distribute information [3]. The information system is a combination of people, hardware, software, communications networks, and data sources compiled, transformed, and experienced streaming processes within an organization [4].

Libraries are institutions managing the collection of paperworks, prints, and / or professional record work with standard systems to meet the educational, research, preservation, information, and recreation needs of the users. The collection of libraries is all information in the form of papers, prints, and / or recording works in various media that have educational value, compiled, processed and served [1]. Each university organizes libraries that meet the national standards of the library regarding the National Education Standards. The college library develops library services based on information and communication technologies. The library as a management system of recording the ideas, thoughts, experiences, and knowledge of mankind, has the main function of preserving the cultural products of mankind, especially in the form of documents of print and other record works, and conveying the ideas, thoughts, experiences, and knowledge of mankind to the next generations. The objective of the implementation of this function is the formation of a society that has a lifelong reading and learning culture. On the other hand, the library serves to support the National Education System as regulated by Law Number 20 Year 2003 on National Education System. The library is a center of information, science, technology, art and culture [5].

An information system audit is the process of collecting and evaluating facts to determine whether an information system protects assets, has data integrity, and helps organizational goals can be achieved [6]. The audit activities of the SLiMS library information system at several universities refer to the ISO 9126 standard. The ISO 9126 standard is used because it has a fairly good compromise in the scope of management and the detail of its processes so it is expected to have clear and integrated guidelines in the use of information technology.

2. Basic Concept of SLiMS

Senayan Library Management System (SLiMS) is a library management system software with open source licensed under GPL v3. This application was first developed and used by the Library of the Ministry of National Education, Information Center and Public Relations, Ministry of National Education. As time goes by, this application was developed by the user community and SLiMS activists. The SLiMS application is built using PHP, MySQL database, and Git version controllers. In 2009, SLiMS received first-rate awards in the 2009 INAICTA event for the open source category.

The following are some of the facilities available to users of the Senayan Library Management System (SLiMS) [2] application, including the Online Public Access Catalog (OPAC) with thumbnails featuring book covers; there is a simple search mode (Simple Search) and Advanced Search; book description details are also available XML format (Extensible Markup Language) for web service needs; efficient bibliographic data management minimizes data redundancy; masterfile / dictionary table management for referential data such as GMD (General Material Designation), Collection Type, Publisher, Author, Location, Keywords and others; circulation with features: lending transactions, returns, collection reservations, flexible lending rules, late information and fines; membership management including the creation of membership cards; inventory management (stocktaking) collection; reports and Statistics; management of periodical publications (Kardex); support management of multimedia documents (.flv, .mp3) and other digital documents (especially for pdfs in streaming form); supports multiple language formats including languages that do not use writing other than Latin; application introductory languages available in Indonesian, English, Spanish, Arabic, German, Thai, Persian; support Union Service Module; visitor counter / library member attendance; login for members from the OPAC page to view the collection members are borrowing; system modules with features: Global system configuration, Module Management, User Management (Library Staff) and groups, Holiday settings, Auto barcode creation, and Utilities for backup; copy cataloging with protocol z39.50, MARC format, and p2p service; and notification of late mail letters by e-mail using mail server.

3. Basic Concept of ISO 9126

Software testing is the process of executing programs intensively to find mistakes. Testing is not only for getting the right program, but also ensuring that the program is free of errors for all conditions [4]. Software testing is a critical element of software quality assurance and presenting specifications, design and coding [7].

ISO 9126 is one of the international standard frameworks used to perform software quality testing, made by the International Organization for Standardization (ISO) and International Electrotechnical Commission (IEC). This international standard has the ability to define the quality of software products, quality characteristics, models, and related metrics to evaluate and define the quality of a software product. The ISO 9126 model has 6 characteristics and several sub-characteristics, as shown in the following table of characteristic characteristics and sub-characteristics of the ISO 9126 model: [8]

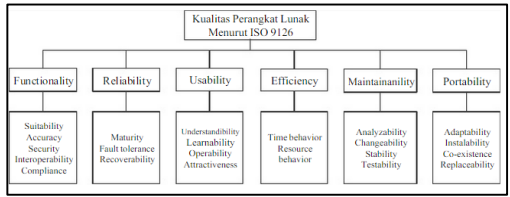

Figure 1. Characteristics and Sub-Characteristics of ISO 9126

Based on Figure 1 above can be explained the six characteristics of the model of ISO 9126, among others, Functionality (Functionality) is the ability of the software in providing functions in accordance with the needs of users when used under specific conditions. A website must be able to be accessed by users with different system environments without compromising existing functionality; Reliability is the ability of software to maintain its performance when used under certain conditions. The reliability of a software can be determined from the number of entries that can cause failure while it is being run. This can be observed by the user; Usability refers to whether a product can achieve a particular goal effectively, efficiently, and gain satisfaction after use. Usability aspects can be measured using a questionnaire instrument. This questionnaire will be filled out by users after they try to use the app. Ease of use is the degree to which software is easy to use, where it is often indicated using sub-attributes for ease of understanding, ease of learning, and operability;

Efficiency relates to the ability of the software to provide a corresponding performance against the amount of resources used in such circumstances. If you want to create a reliable software, the efficiency aspect should be really taken care of. Inefficient use of resources, for example, using improper algorithms can cause software performance to be sluggish; Maintainability relates to the ability of the software to be modified. Modifications include correction, improvement or adaptation to environmental changes, requirements, and functional specifications; and Portability (Portability) relates to the ability of a software to operate or work in different environments. To test the portability level of a web-based application, the application is attempted to run using a different browser. In addition, experiments using different devices can also determine the portability level of the software

4. Related Research

A study has been conducted to make adjustments to software quality models that fit the characteristics of business applications. The result of this research is that software in business activity has some unique characteristics. The general ISO 9126 is perceived to be incapable of covering all the unique characteristics of a particular software, one of which is a business application. It is found the main characteristics in a business application that also involves the developer, namely Usability, Portability, Efficiency, Reliability, Functionality, and Traceability [9].

Other research has been done in making the quality standards of game apps on mobile devices so that developers can have a reference to assess game software to be developed. The result of this study proves that with reference to ISO 9126 developers can pay more attention and consider factors that have a greater significance value in this case is the aspect of functionality, usability, and portability [10].

The previous research has been done related to the audit problem of Axapta ERP system in PT Posmi Steel Indonesia using ISO 9126 so that the company has a reference in the development of existing ERP system. The results of the research indicate that additional customization facilities are required on the forms and reports contained in the Axapta ERP module so that they are able to generate reports according to the schedule determined by the management [11].

Other studies have also been conducted to determine the level of community satisfaction with services performed by Bitungsari Bogor administrative apparatus related to administration and other activities. Data analysis techniques used in the study are questionnaires and data processing methods using the Satisfaction Index Society in accordance with KEP / 25 / M.PAN / 2/2004. Based on the results of data processing according to KEP / 25 / M.PAN / 2/2004 it can be seen that overall that the level of service of state apparatus in Bitungsari sub-district is at the level of GOOD where the certainty of service cost and environmental comfort get VERY GOOD appreciation while speed of service gets spotlight should be noticed even if it is still at a GOOD level [12].

## C. Research Methodology

The research was conducted by using descriptive quantitative method. The collection of data and information is closely related to user requirement using the application of Senayan Library Management System (SLiMS) in the Library of Higher Education. Data and information obtained from primary data and secondary data. Primary data was obtained by distributing questionnaires about the application of Senayan Library Management System (SLiMS) at Indonesian Universities Library, while secondary data was obtained through literature study, through literature study and scientific writing on Library Information System and ISO 9126.

For primary data, obtained from the questionnaire with respondents as many as 10 (ten) respondents who are librarians in charge from IPMI IBS, Bakrie University, Perbanas Institute Jakarta, STMIK Bina Insani, Prasetya Mulya University, Agung Podomoro University, Indonesian Law High School of Jentera, Matana University, STIKS Tarakanita Jakarta, and STAI-PIQ West Sumatra. Respondents fill out their opinions regarding the experience related to the use of SLiMS applications in their respective college libraries.

In this research, the measured variable is the result of the user experience of SLiMS on the ISO 9126 information system audit variable using Functionality, Reliability, Usability, Efficiency, Maintainability and Portability. Each aspect uses a Likert scale to measure attitudes, opinions, and perceptions of respondents. With the Likert scale, the variable to be measured is translated into a variable indicator. Then the indicator is used as a starting point to arrange items in the form of questions or statements (Sugiyono, 2008: 93). Questionnaire filling has been done by the librarians in charge as well as direct users of SLiMS applications. The answer available is strongly agree with the value 4, agree with the value 3, disagree with the value 2, strongly disagree with the value 1.

There are about 21 items statement representing each aspect Functionality, Reliability, Usability, Efficiency, Maintainability, and Portability.

After the data is complete then the next step is data processing by counting the number of respondents who strongly agree, agree, disagree, and strongly disagree and make Likert scale for the total of all attributes and each attribute is Functionality, Reliability, Usability, Efficiency, Maintainability, and Portability. Likert scale is made with the formula total answer x value x number of respondents. Value for very good is 4, good is 3, bad is 2, and very bad is 1. So, Likert scale for very good is the number of answers x 4 x questions, for good is the number of answers x 3 x questions, for bad is the number of answers x 2 x questions, and for very bad is the number of answers x 1 x questions.

**D. Result and Discussion**
1. Results of Data Analysis Based on All Attributes in ISO 9126

The basis used to audit the SLiMS information system of all attributes as a whole is the Likert Scale with the following calculations:

a. Very good = 10 respondents x 4 x 21 questions = 840

b. Good = 10 respondents x 3 x 21 questions = 630

c. Bad = 10 respondents x 2 x 21 questions = 420

d. Very bad = 10 respondents x 1 x 21 questions = 210

The results of respondents' opinions obtained by summing all the answers for all attributes, obtained the total value of the answer is 728. By comparing the results of respondents and Likert scale, then the value 728 is classified as very good. This means that overall SLiMS application is very good to be used as library management in IPMI IBS, Bakrie University, Perbanas Institute Jakarta, STMIK Bina Insani, Prasetya Mulya University, Agung Podomoro University, Indonesian Law High School of Jentera, Matana University, STIKS Tarakanita Jakarta, and STAI-PIQ of West Sumatra.

4.2. Result of Data Analysis Based on Attribute Functionality

The basis used to audit the SLiMS information system of the Functionality attribute is the Likert Scale with the calculation:

a. Very good = 10 respondents x 4 x 4 questions = 160

b. Good = 10 respondents x 3 x 4 questions = 120

c. Bad = 10 respondents x 2 x 4 questions = 80

d. Very bad = 10 respondents x 1 x 4 questions = 40

The result of the respondent's opinion obtained by summing all the answers for the Functionality attribute, obtained the total answer value is 148. By comparing the results of the respondent's answer and Likert scale, then the value of 148 goes into very good classification. This means that SLiMS applications provide functionality according to the needs of users when used in specific conditions so as to be accessible to users with different system environment without reducing the existing functionality.

4.3. Result of Data Analysis Based on Reliability Attribute

The basis for the SLiMS information system audit of the Reliability attribute is the Likert Scale with the calculation:

a. Very good = 10 respondents x 4 x 3 questions = 120

b. Good = 10 respondents x 3 x 3 questions = 90

c. Bad = 10 respondents x 2 x 3 questions = 60

d. Very bad = 10 respondents x 1 x 3 questions = 30

Reliability is the ability of software to maintain its performance when used under certain conditions. The reliability of a software can be determined from the number of entries that can cause failure while it is being run. This can be observed by the user. The result of the respondent's opinion obtained by summing all the answers for the Reliability attribute, obtained the total value of the answer is 105. By comparing the results of the respondent's answer and Likert scale, then the value of 105 goes into good classification. This means that SLiMS applications are excellent in terms of maintaining their performance when used under certain conditions, where the reliability of SLiMS applications can be determined from the number of entries that can cause failure while it is being run.

## 4.4. Result of Data Analysis Based on Usability Attribute

The basis used for audit of SLiMS information system from Usability attribute is Likert Scale with calculation:

a. Very good = 10 respondents x 4 x 4 questions = 160
b. Good = 10 respondents x 3 x 4 questions = 120
c. Bad = 10 respondents x 2 x 4 questions = 80
d. Very bad = 10 respondents x 1 x 4 questions = 40

The result of the respondent's opinion obtained by summing all the answers for the Usability attribute, obtained the total answer value is 130. By comparing the results of the respondent's answer and Likert scale, then the value of 130 goes into very good classification. This means that SLiMS applications are easy to understand, studied and operated by the librarian in charge.

## 4.5. Result of Data Analysis Based on Attribute Efficiency

The basis used to audit the SLiMS information system of the Efficiency attribute is the Likert Scale with the following calculations:

a. Very good = 10 respondents x 4 x 2 questions = 80
b. Good = 10 respondents x 3 x 2 questions = 60
c. Bad = 10 respondents x 2 x 2 questions = 40
d. Very bad = 10 respondents x 1 x 2 questions = 20

The results of respondents' opinions obtained by summing all the answers for the Efficiency attribute, obtained the total value of the answer is 67. By comparing the results of respondents and Likert scale, then the value 67 is classified as very good. This means that the SLiMS application provides the appropriate performance against the amount of resources used in those circumstances.

## 4.6. Result of Data Analysis Based on Maintainability Attribute

The basis used to audit the SLiMS information system of the Maintainability attribute is the Likert Scale with the calculation:

a. Very good = 10 respondents x 4 x 4 questions = 160

b. Good = 10 respondents x 3 x 4 questions = 120

c. Bad = 10 respondents x 2 x 4 questions = 80

d. Very bad = 10 respondents x 1 x 4 questions = 40

The results of respondents' opinions obtained by summing all the answers to the Maintainability attribute, obtained the total value of the answer is 143. By comparing the results of respondents and Likert scale, then the value of 143 is classified as very good. This means that SLiMS applications have the ability to be modified which includes correction, repair or adaptation to environmental changes, requirements, and functional specifications.

4.7. Result of Data Analysis Based on Portability Attribute

The basis used for audit of SLiMS information system from Portability attribute is Likert Scale with calculation:

a. Very good = 10 respondents x 4 x 4 questions = 160

b. Good = 10 respondents x 3 x 4 questions = 120

c. Bad = 10 respondents x 2 x 4 questions = 80

d. Very bad = 10 respondents x 1 x 4 questions = 40

The results of respondents' opinions obtained by summing all the answers to the Portability attribute, obtained the total answer value is 135. By comparing the results of respondents and Likert scale, the value of 135 is classified as very good. This means that SLiMS applications are able to operate or work in different environments. It is proven that SLiMS applications are able to operate or work very well in the existing libraries of IPMI IBS, Bakrie University, Perbanas Institute Jakarta, STMIK Bina Insani, Prasetya Mulya University, Agung Podomoro University, Indonesian Law School of Jentera, University of Matana, STIKS Tarakanita Jakarta, and STAI-PIQ West Sumatra.

**E. Conclusion**

1. Conclusion

Based on the results of the audit information system using ISO 9126 against SLiMS that SLiMS is classified as the very well category. In other words that SLiMS has a familiar look for the librarian in charge so it does not take a long time to understand its use. SLiMS also proved easy to integrate data with other file form, especially Ms. Excel. This means that SLiMS is very useful to use in the management of libraries in ten universities where the research was conducted, namely IPMI IBS, Bakrie University, Perbanas Institute Jakarta, STMIK & Bina Insani Academy, Prasetya Mulya University, Agung Podomoro

University, Indonesian Law High School of Jentera, Matana University, STIKS Tarakanita Jakarta, and STAI-PIQ West Sumatra. SLiMS can be viewed very well in aspects of Functionality, Reliability, Usability, Efficiency, Maintainability, and Portability.

2. Suggestion

For further research, it should be able to use the latest framework that is ISO 25010 to perform software quality audits. The number of respondents should also be cultivated more widely spread throughout the territory of Indonesia. As for the distribution of questionnaires to reach a broad range should use the management of an online questionnaire, such as google form, monkey surveys, and so forth.